\documentclass{pasj01}

\RequirePackage{mathpazo}
\RequirePackage[T1]{fontenc}

\begin{document} 

\SetRunningHead{A.Imada et al.}{}

\Received{2017/01/27}
\Accepted{2017/06/05}

\title{The 2015 superoutburst of QZ Virginis: Detection of growing superhumps between the precursor and main superoutburst}

\author{Akira \textsc{Imada}\altaffilmark{1}}
\altaffiltext{1}{Kwasan and Hida Observatories, Kyoto University, Yamashina, Kyoto 607-8471, Japan}
\email{a\_imada@kwasan.kyoto-u.ac.jp}

\author{Taichi \textsc{Kato}\altaffilmark{2}}
\author{Keisuke \textsc{Isogai}\altaffilmark{2}}
\altaffiltext{2}{Department of Astronomy,Faculty of Science, Kyoto University,
      Sakyo-ku, Kyoto 606-8502, Japan}

\author{Franz-Josef \textsc{Hambsch}\altaffilmark{3}\altaffilmark{4}\altaffilmark{5}}
\altaffiltext{3}{Groupe Europ\'een d'Observations Stellaires (GEOS), 23 Parc de Levesville, 28300 Bailleau l'Ev\^eque, France}
\altaffiltext{4}{Bundesdeutsche Arbeitsgemeinschaft f\"ur Ver\"anderliche Sterne (BAV), Munsterdamm 90, 12169 Berlin, Germany}
\altaffiltext{5}{Vereniging Voor Sterrenkunde (VVS), Oude Bleken 12, 2400 Mol, Belgium}

\author{Pavol A. \textsc{Dubovsky}\altaffilmark{6}}
\author{Igor \textsc{Kudzej}\altaffilmark{6}}
\altaffiltext{6}{Vihorlat Observatory, Mierova 4, 06601 Humenne, Slovakia}

\author{Roger D. \textsc{Pickard}\altaffilmark{7}}
\altaffiltext{7}{The British Astronomical Association, Variable Star Section (BAA VSS), Burlington House, Piccadilly, London, W1J 0DU, UK; 3 The Birches, Shobdon, Leominster, Herefordshire, HR6 9NG, UK}

\author{Hidehiko \textsc{Akazawa}\altaffilmark{8}}
\altaffiltext{8}{Department of Biosphere-Geosphere System Science, Faculty of Informatics, Okayama University of Science, 1-1 Ridai-cho, Okayama, Okayama 700-0005, Japan}

\author{Kiyoshi \textsc{Kasai}\altaffilmark{9}}
\altaffiltext{9}{Baselstrasse 133D, CH-4132 Muttenz, Switzerland}

\author{Hiroshi \textsc{Itoh}\altaffilmark{10}}
\altaffiltext{10}{Variable Star Observers League in Japan (VSOLJ), 1001-105 Nishiterakata, Hachioji, Tokyo 192-0153, Japan}

\author{Lewis M. \textsc{Cook}\altaffilmark{11}}
\altaffiltext{11}{Center for Backyard Astrophysics Concord, 1730 Helix Ct. Concord, CA 94518, USA}

\author{Seiichiro \textsc{Kiyota}\altaffilmark{12}}
\altaffiltext{12}{VSOLJ, 7-1 Kitahatsutomi, Kamagaya, Chiba 273-0126, Japan}


\maketitle

\begin{abstract}

We report on time-resolved photometry of the 2015 February-March superoutburst of QZ Virginis. The superoutburst consisted of a separated precursor, main superoutburst, and rebrightening. We detected superhumps with a period of 0.061181(42) d between the precursor and main superoutburst. Based on analyses of period changes and amplitudes of superhumps, the observed superhumps were identified as growing superhumps (stage A superhumps). The duration of stage A superhumps was about 5 d, unusually long for SU UMa-type dwarf novae. Using the obtained stage A superhump period, we estimated the mass ratio of QZ Vir to be 0.108(3). This value suggests that QZ Vir is an SU UMa-type dwarf nova evolving toward the period minimum. Based on the present and the previous observations regarding long-lasting stage A superhumps, a time scale of stage A superhumps is likely to be determined by the mass ratio of the system and the temperature of the accretion disk.
\end{abstract}

\section{Introduction}

Cataclysmic variables are close binary systems that consist of a primary white dwarf and a secondary star. The secondary star fills its Roche lobe, transferring mass into the primary Roche lobe. If the magnetic field of the white dwarf is weak ($<$ 10 MG), an accretion disk is formed around the white dwarf (for a review, see \cite{war95book}; \cite{hel01book}). 

Dwarf novae are a subclass of cataclysmic variables (for a review, see \cite{osa96review}). A long-term light curve of dwarf novae is well reproduced by the disk instability model \citep{osa74DNmodel}. When the surface density of the accretion disk reaches the high critical point, an instability within the accretion disk sets in and a large amount of the gas fall onto the white dwarf. This is observed as an outburst. On the other hand, when the surface density reaches the low critical point during the outburst, the gas accretion ceases. This is observed as the end of the outburst. The physical mechanism for the disk instability model is well explained by the thermal limit cycle instability (thermal instability) of ionized and neutral hydrogen (\cite{mey81DNoutburst}; \cite{can93DIreview}; \cite{las01DIDNXT}). 

SU UMa-type dwarf novae, one subclass of dwarf novae, have two types of outbursts. One is the normal outburst, whose duration is a few days. The other is the superoutburst, whose duration is as long as 10 days and the maximum magnitude is brighter than that of the normal outburst. The light curve of the normal outburst is reproduced by the thermal instability model. On the other hand, the light curve of the superoutburst is reproduced by the combination of the thermal and tidal instability model \citep{osa89suuma}.\footnote{On the other hand, there are two models in order to explain the superoutburst: the enhanced mass transfer model in which the superoutburst is caused by the enhanced mass transfer from the secondary (e.g., \cite{sma91suumamodel}; \cite{sch04vwhyimodel}), and the pure thermal limit cycle model in which the superoutburst is explained by the thermal limit cycle model (e.g., \cite{can10v344lyr}; \cite{can12v344lyr}). See \citet{osa13v1504cygKepler} for a review.} When the disk radius reaches the 3:1 resonance between the Keplerian motion of the disk and the orbital motion of the secondary, the gas at the 3:1 resonance radius deviates from the circular orbit, which propagates into the entire disk. As a result, the accretion disk is deformed to a non-axisymmetric structure, which causes a significantly enhanced tidal dissipation and torque on the accretion disk (\cite{whi88tidal}; \cite{hir90SHexcess}). This results in a long-lasting outburst (superoutburst). During the superoutburst, tooth-like modulations with amplitudes of ${\sim}$ 0.2 mag, termed superhumps, are observed. The superhump period is slightly longer (typically 1${\sim}$3 \% longer) than the orbital period of the system. This is understood as prograde precession of a tidally-deformed eccentric accretion disk \citep{osa89suuma}.

One of the most important studies concerning photometry of superoutbursts is to examine how the superhump period changes during the course of the superoutburst. Extensive statistical surveys for the superhump period changes have been performed by T. Kato and his colleagues (\cite{Pdot}; \cite{Pdot2}; \cite{Pdot3}; \cite{Pdot4}; \cite{Pdot5}; \cite{Pdot6}; \cite{Pdot7}; \cite{Pdot8}). They have established a "textbook'' of superhump period changes, which consist of three stages: a long and constant period (stage A), increasing period (stage B), and short and constant period (stage C) (see also figure 3 of \citet{Pdot}). Although the working mechanisms causing superhump period changes are still unclear, our understanding on stage A superhumps has been significantly improved over the past few years. 

\citet{osa13v344lyrv1504cyg} studied Kepler light curves of SU UMa-type dwarf novae V344 Lyr and V1504 Cyg. They suggest that superhump period changes are caused by dynamical precession, pressure effect, or wave-wave interaction of the accretion disk. \citet{osa13v344lyrv1504cyg} further noted that pressure effect and wave-wave interaction are negligible at the onset of appearance of superhumps. This means that the period of stage A superhumps corresponds to the dynamical precession rate of the accretion disk at the 3:1 resonance radius. This interpretation is supported by \citet{kat13qfromstageA}, in which they studied eclipsing dwarf novae and showed that the mass ratios derived by stage A superhump periods are in good agreement with those derived by eclipses. At present, measuring stage A superhump period has become one of the most powerful tools for estimating the mass ratio of the system (\cite{kat14j0902}; \cite{ohs14eruma}; \cite{iso16crboo}) .

Statistical studies have revealed that the system showing a long duration of stage A superhumps tends to have a small mass ratio \citep{kat15wzsge}. This trend is particularily seen in WZ Sge-type dwarf novae (for a review of WZ Sge-type dwarf novae, see \cite{kat15wzsge}). In recent years, observations showed that stage A superhumps lasted for ${\sim}$ 5 d in a faint stage between the precursor and the main superoutburst of a newly confirmed SU UMa-type dwarf novae PM J03338+3320 \citep{kat16j0333}. Although the duration of stage A superhumps in PM J03338+3320 was comparable to that of WZ Sge-type dwarf novae, the mass ratio was estimated to be 0.17, an average value for SU UMa-type dwarf novae \citep{kat16j0333}. At present, although it is a very rare event, two SU UMa-type dwarf novae showed long-lasting stage A superhumps in a faint stage between the precursor and the main superoutburst: PM J03338+3320 and V1504 Cyg (\cite{kat16j0333}; \cite{osa14v1504cygv344lyrpaper3}).

QZ Vir (originally named as T Leo) is one of the most notable dwarf novae. The observational history of the object dates back to the 1860s \citep{pet65tleo}. Spectroscopic observations revealed the short orbital period of 84.69936(68) min (= 0.058819(1) d) \citep{sha84tleo}. The SU UMa nature of the object was confirmed by \citet{iauc4314} and \citet{kat87tleo} during the 1987 January superoutburst after detection of superhumps. However, the lack of observations prevented an accurate determination of the superhump period. \citet{lem93tleo} performed photometry of QZ Vir during the 1993 January superoutburst, during which they obtained the mean superhump period of 86.7 $\pm$ 0.1 min (0.060208(69) d). This period was also confirmed by \citet{kat97tleo} in the same superoutburst. On the other hand, some authors have pointed out unusual properties of QZ Vir. For example, \citet{vri04tleo} detected a signal at 414 s in X-ray light curves, which they interpreted as the spin period of the white dwarf, or quasi-periodic oscillations. \citet{sha84tleo} pointed out a possibility of unusually large mass ratio of the system ($q {\sim}$ 0.48, see also figure 11 of \cite{sha84tleo}), based on their spectroscopic and photometric observations during quiescence. These arguments have obscured an evolutional status of QZ Vir. 

On 2015 February 21.512 (JD 2457075.012), Rod Stubbings reported that QZ Vir was at a visual magnitude of 15.0, about 1 mag brighter than quiescence. After that, QZ Vir brightened to 13.3 on February 21.692 ([vsnet-alert 18318]). On 2015 February 22.919, a visual magnitude reached 10.6, then the system faded again ([vsnet-alert 18334]). On 2015 March 4.885 (JD 2457086.385), QZ Vir rebrightened at a visual magnitude of 10.5 ([vsnet-alert 18371]). It turned out that the latter outburst was indeed a superoutburst which was accompanied by a separate precursor, similar to that observed in PM J03338+3320 and V1504 Cyg. Thanks to the early discovery of the outburst, we succeeded in detecting growing superhumps (stage A superhumps) of QZ Vir from the end of the precursor in unprecedented detail. The present observations enables us to determine the evolutional status of the system for the first time, and improves our understanding of stage A superhumps. In section 2, we present our observations. In section 3, we show the main results of the superoutburst. Discussion is given in section 4. We summarize our studies in section 5.

\begin{longtable}{lllllll}
\caption{Log of Observations.}\label{tlog}
\hline\hline
Date & JD(start)$^*$ & JD(end)$^*$ & N$^{\dagger}$ & Filter$^{\ddagger}$ & Code${\S}$ & Exp(sec)$^{||}$\\
\hline
\endhead
\endfoot
2015 Feb. 26 & 7080.4585 & 7080.5539 & 108 & V & RPc  & 60 \\
2015 Feb. 27 & 7080.6397 & 7080.8782 & 100 & C & HaC  & 180 \\
2015 Mar. 2 & 7083.6313 & 7083.8784 & 106 & C & HaC  & 180 \\
2015 Mar. 3 & 7084.6288 & 7084.8809 & 104 & C & HaC  & 180 \\
2015 Mar. 4 & 7085.6266 & 7085.8801 & 107 & C & HaC  & 180 \\
            & 7086.4471 & 7086.4513 & 12 & V & RPc & 30 \\
2015 Mar. 5 & 7086.6238 & 7086.6312 & 4 & C & HaC & 180 \\
2015 Mar. 7 & 7089.3890 & 7089.5928 & 248 & V & DPV  & 60 \\
2015 Mar. 9 & 7090.6127 & 7090.8884 & 120 & C & HaC  & 180 \\
            & 7091.0860 & 7091.3328 & 226 & R & Aka  & 90 \\
            & 7091.1071 & 7091.2235 & 105 & V & Aka  & 90 \\
            & 7091.4317 & 7091.5975 & 214 & V & DPV  & 60 \\
2015 Mar. 10& 7091.7487 & 7091.8891 & 118 & C & HaC  & 180 \\
            & 7092.1508 & 7092.3073 & 146 & R & Aka  & 90 \\
            & 7092.1670 & 7092.3321 & 157 & V & Aka  & 90 \\
2015 Mar. 11& 7092.6076 & 7092.8901 & 193 & C & HaC  & 180 \\
            & 7093.0797 & 7093.1483 & 65 & R & Aka  & 90 \\
            & 7093.1042 & 7093.1497 & 44 & V & Aka  & 90 \\
            & 7093.1327 & 7093.1624 & 79 & C & Kis  & 30 \\
2015 Mar. 12& 7093.6048 & 7093.8893 & 70 & C & HaC  & 180 \\
            & 7094.1099 & 7094.1753 & 62 & R & Aka  & 90 \\
            & 7094.1173 & 7094.1797 & 58 & V & Aka  & 90 \\
2015 Mar. 13& 7094.6021 & 7094.8886 & 114 & C & HaC & 180 \\
2015 Mar. 14& 7095.5086 & 7095.6399 & 51 & C & COO  & 120 \\
            & 7095.5999 & 7095.8495 & 90 & C & HaC  & 180 \\
            & 7096.0878 & 7096.2277 & 162 & C & Ioh  & 40 \\
            & 7096.4042 & 7096.4723 & 89 & C & Kai & 60 \\ 
2015 Mar. 15& 7096.6036 & 7096.8457 & 66 & C & HaC  & 180 \\
2015 Mar. 16& 7097.5943 & 7097.8427 & 77 & C & HaC  & 180 \\
            & 7098.3317 & 7098.5606 & 154 & C & DPV  & 120 \\
2015 Mar. 17& 7099.3796 & 7099.6137 & 156 & C & DPV  & 120 \\
2015 Mar. 21& 7102.6837 & 7102.8313 & 50 & C & HaC  & 180 \\
2015 Mar. 23& 7104.7315 & 7104.8258 & 44 & C & HaC  & 180 \\
            & 7105.2813 & 7105.5438 & 177 & C & DPV  & 120 \\
            & 7105.3643 & 7105.5131 & 204 & C & Kai  & 60 \\
2015 Mar. 24& 7106.3786 & 7106.4538 & 45 & C & DPV  & 120 \\
2015 Mar. 27& 7108.5895 & 7108.8136 & 73 & C & HaC  & 180 \\
2015 Mar. 28& 7109.5772 & 7109.8068 & 116 & C & HaC  & 180 \\
2015 Mar. 29& 7110.5679 & 7110.8093 & 120 & C & HaC  & 180 \\
2015 Mar. 30& 7111.5696 & 7111.8066 & 108 & C & HaC  & 180 \\
2015 Apr.  6& 7119.3805 & 7119.5501 & 216 & C & RPc & 60 \\
2015 Apr.  7& 7120.3883 & 7120.4845 & 59 & C & RPc & 120 \\
2015 Apr.  9& 7122.3924 & 7122.4653 & 72 & C & RPc & 60 \\
\hline
\multicolumn{7}{l}{$^*$ JD-2450000. $^{\dagger}$ Numbers of images.} \\
\multicolumn{7}{l}{$^{\ddagger}$ The filter name $C$ represents unfiltered observations.} \\
\multicolumn{7}{l}{$^{\S}$ Code of observers: RPc (R. Pickard, 34cm), HaC (J. Hambsch, 40cm), DPV (P. Dubovsky, 28cm)} \\
\multicolumn{7}{l}{Aka (H. Akazawa, 28cm for $V$ filter, 20cm for $R$ filter), Kis (S. Kiyota, 25cm)} \\
\multicolumn{7}{l}{COO (L. Cook, 32cm), Ioh (H. Itoh, 30cm), Kai (K. Kasai, 28cm).} \\
\multicolumn{7}{l}{$^{||}$ Exposure time in unit of seconds.} \\
\end{longtable}

\begin{figure*}
\begin{center}
\includegraphics[width=16cm]{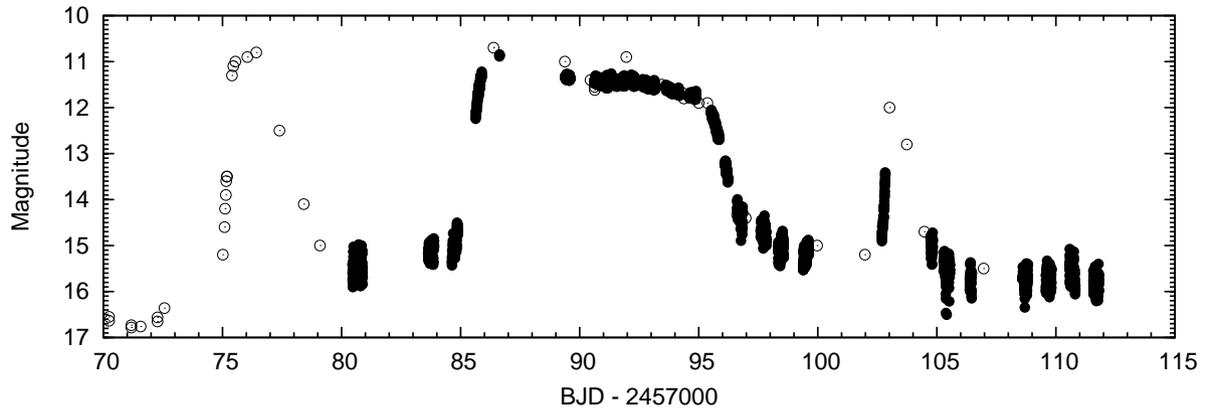}
\end{center}
\caption{Overall light curve during the 2015 superoutburst. Black points indicate CCD observations, while white points indicates visual observations posted to the AAVSO. Note that the main superoutburst is accompanied by a separated precursor and rebrightening.}
\label{lc}
\end{figure*}

\section{Observations}

Time-resolved CCD photometry was performed between 2015 Feb. 26 and 2015 Mar. 30(JD 2457080.4585$-$7111.8066) at 8 sites using 20$-$40 cm reflecting telescopes. The log of observations is listed in table \ref{tlog}. Although some of the data were acquired with $V$ and $R$ band filters, the most of the data were acquired without filters. The total datapoints amount to 4689, sufficient to study superhump period changes. Exposure times were 30$-$180 seconds, with read-out times typically an order of seconds.

After debiasing and flat-fielding, the images were processed with aperture photometry (see table \ref{tlog}). The data obtained were adjusted to the HaC system, in which the star located at RA:11:37:55.92 Dec:03:23:13.5 (V = 13.247, R = 12.847) was used as the comparison star. The constancy of the comparison star was checked mainly by the star located at RA:11:38:30.54 Dec:03:25:53.3 (V = 14.396, R = 13.970) and nearby stars in the same image. The times of all observations were converted to Barycentric Julian Date (BJD).

\section{Results}

\begin{figure}
\begin{center}
\includegraphics[width=8cm]{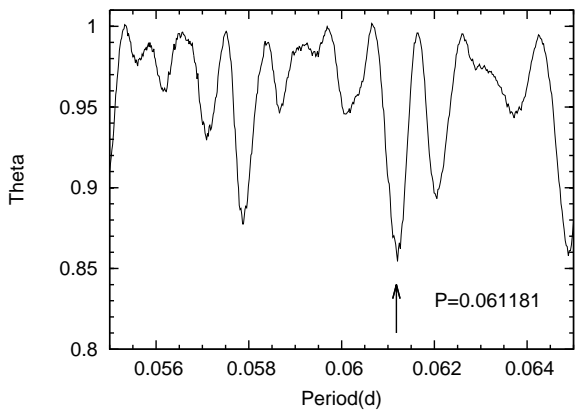}
\end{center}
\caption{PDM analysis between BJD 2457080.64$-$85.88. The strongest signal corresponds to 0.061181(42) d, 4.0 \% longer than the orbital period of QZ Vir.}
\label{pdmpre}
\end{figure}

\begin{figure}
\begin{center}
\includegraphics{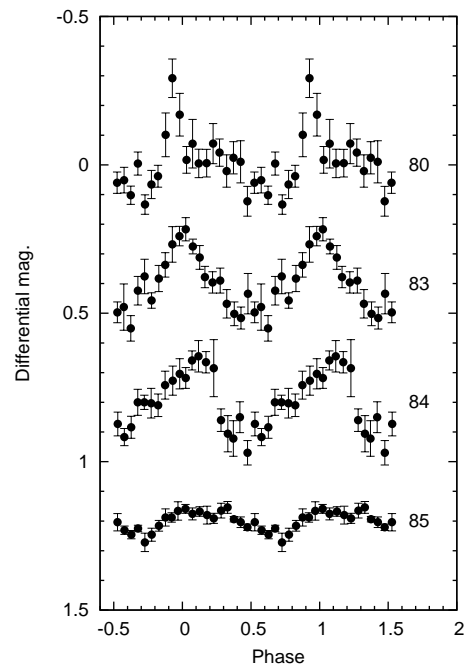}
\end{center}
\caption{Phase-averaged light curves folded with 0.061181 d. The numbers in this figure denote the day since BJD 2457000.}
\label{shvar1}
\end{figure}

\subsection{Overall light curve}

Figure \ref{lc} shows the overall light curve of the present outburst. As can be seen in this figure, the light curve consists of three outbursts. Although the initial outburst is separated from the subsequent superoutburst, this outburst can be regarded as a precursor, since superhumps were detected after the end of the precursor (see also the next subsection). The duration of the faint stage between the precursor and the main superoutburst was ${\sim}$ 5 d. We caught the rising stage of the superoutburst on BJD 2457085, when QZ Vir brightened at a rate of steeper than $-$3.3 mag/d. This value was typical for SU UMa-type dwarf novae \citep{otu16dn}. Although our observations were absent between BJD 2457086$-$88, the brightness maximum may have occurred around BJD 2457086. The magnitude kept almost constant at ${\sim}$ 11.5 mag between BJD 2457089$-$91, after which the magnitude linearly faded at a rate of 0.13 mag/d. The duration of the plateau stage was ${\sim}$ 9 d. On BJD 2457095, QZ Vir entered a rapid fading stage, during which the magnitude faded at a rate of 1.9 mag/d. After the end of the main superoutburst, the magnitude slowly faded at a rate of 0.2 mag/d. A rebrightening was observed on BJD 2457103. The light curve of the rebrightening was typical for that observed in SU UMa-type dwarf novae (see, e.g., figure 55 of \cite{Pdot3}, figure 16 of \cite{Pdot5}).

\subsection{Superhumps}

One of the noticable results was that superhumps appeared from the end of the precursor. On BJD 2457080, when the magnitude returned to the quiescent level, superhumps were visible with an amplitude of ${\sim}$ 0.8 mag. On BJD 2457083$-$84, the light curve still sustained superhumps with amplitudes of ${\sim}$ 0.4 mag. During the rising stage of the superoutburst (BJD 2458085), superhumps were barely observed. We performed the phase dispersion minimization method (PDM, \cite{pdm}) for the residual light curve between BJD 2457080$-$85. The 1 sigma errors for the PDM analysis were calculated by the method developed by \citet{fer89error} and \citet{Pdot2}. Figure \ref{pdmpre} illustrates the resultant theta diagram, in which the strongest periodicity corresponds to $P$ = 0.061181(42) d. This period is 4.0 \% longer than the orbital period of QZ Vir, which safely excludes the orbital origin. Figure \ref{shvar1} shows phase-averaged light curves folded by $P$ = 0.061181 d, in which single-peaked profiles are visible. Such a sigle-peaked hump was also observed in an SU UMa-type dwarf nova PM J03338+3320 in the faint stage between the precursor and the main superoutburst \citep{kat16j0333}. They identified the observed single-peaked humps as superhumps, based on the period analyses and the profile of the humps. In combination with the above period, the modulations of the single-peaked light curves, and the previous report by \citet{kat16j0333}, we can reasonably conclude that the single-peaked modulations are indeed superhumps.

In order to examine the characteristics of the superhumps, we calibrated superhump maxima and the amplitudes of superhumps. In general, measuring superhump maxima provide diagnostics of the dynamics in the outbursting accretion disk \citep{Pdot}. The detailed method of calibration is the same as that described in \citet{Pdot}. In table \ref{toc}, we tabulate the maximum timings of the superhumps and $O - C$ values, where we used the following linear regression,

\begin{eqnarray}
BJD(max) = 2457080.5836 + 0.06020 \times E.
\label{eq1}
\end{eqnarray}

\begin{figure}
\begin{center}
\includegraphics[width=8cm]{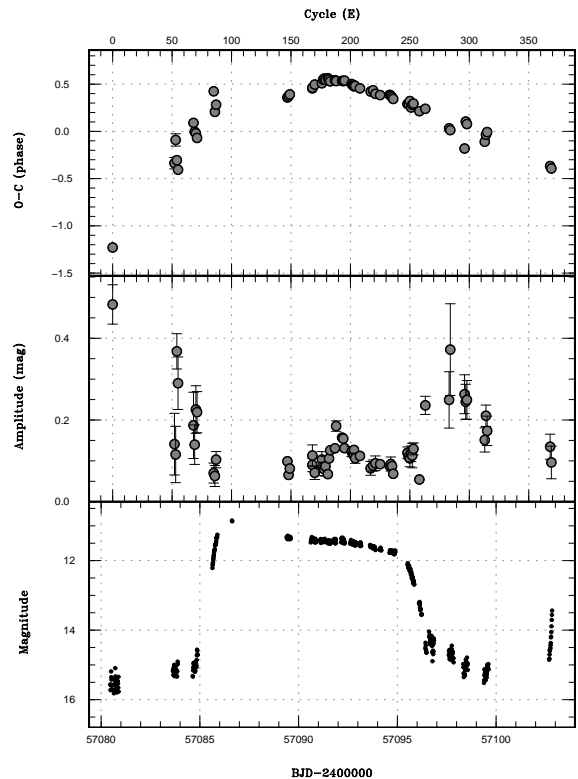}
\end{center}
\caption{$O - C$ diagram of superhump maxima, amplitudes of superhumps, and corresponding light curve. A linear regression is expressed as equation \ref{eq1}. A 'break' is observed on BJD 2457092, when the stage B$-$C transition occurred. Stage A$-$B transition may have occurred between BJD 2457086$-$88, based on the PDM analyses for each segment (see figure \ref{pdm2}).}
\label{o-c}
\end{figure}

\begin{figure}
\begin{center}
\includegraphics[width=8cm]{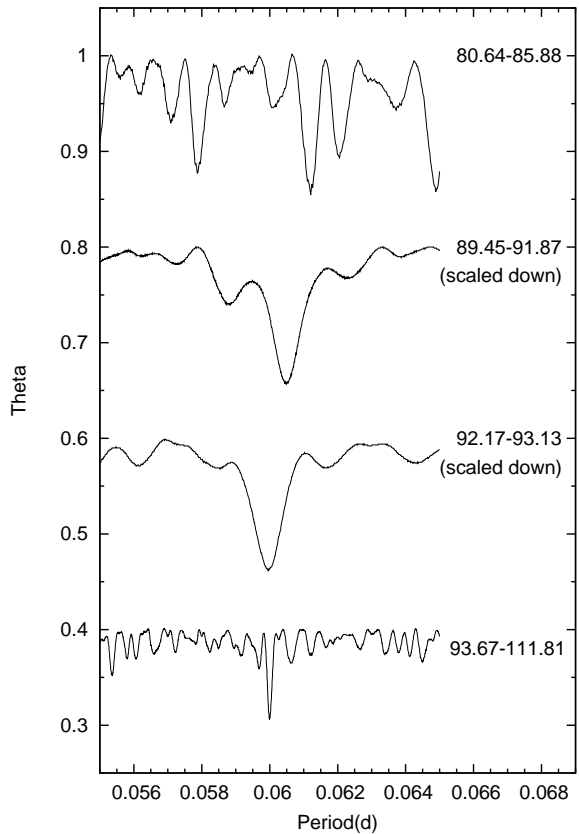}
\end{center}
\caption{Results of PDM analyses. The strongest periodicity corresponds to $P$ = 0.060490(34) d for BJD 2457089.45$-$91.87, $P$ = 0.059944(24) d for BJD 2457092.17$-$93.13, and $P$ = 0.060000(9) d for BJD 2457093.67$-$111.81, respectively. For better visualization, the theta diagrams on BJD 2457089.45$-$91.87 and BJD 2457092.17$-$93.13 are scaled down arbitrarily.}
\label{pdm2}
\end{figure}

Figure \ref{o-c} shows the $O - C$ diagram and amplitudes of the superhumps. As can be seen in these figures, a ``break'' occurred around $E$ ${\sim}$ 180 and $E$ ${\sim}$ 195. Although our photometry was absent near the brightness maximum, the break in the $O - C$ diagram may also have occurred around $E$ ${\sim}$ 100. Based on the obtained $O - C$ diagram, we divided light curves into three segments, BJD 2457089.45$-$91.87 (147$<= E <=$188), BJD 2457092.17$-$93.13 (193$<= E <=$ 208), and BJD 2457093.67$-$111.81 ($E >=$217)\footnote{Because of low quality of the data for period analyses, we did not use the data between BJD 2457119$-$22.}. For each segment, we performed the PDM method and the results of the period analyses are shown in figure \ref{pdm2}. The best estimated period during 147$<= E <=$188 was 0.060490(34) d, which is in good agreement with that obtained in the previous superoutbursts (e.g., $P_{\rm sh}$ = 0.060488 d for the 2005 superoutburst, $P_{\rm sh}$ = 0.060481 d for the 2007 superoutburst, see also table 2 of \citep{Pdot}). We derived $P_{\rm sh}
$ = 0.059944(24) d during 193$<= E <= $208. This value is again in good agreement with that obtained in the previous superoutbursts of stage C superhumps \citep{Pdot}. After the end of the main plateau stage (after BJD 2457095), we derived $P_{\rm sh}$ = 0.060000(9) d, 2.0 \% longer than the orbital period of QZ Vir. This indicates that superhumps were sustained after the end of the main superoutburst.

\section{Discussion}

\subsection{Stage A superhumps in the faint stage between the precursor and the main superoutburst}

As noted above, one of the most important findings is that superhumps appeared after the end of the precursor and survived in the faint stage between the precursor and the main superoutburst. \citet{kat16j0333} recently reported that stage A superhumps were observed in the faint stage of a newly confirmed SU UMa-type dwarf nova PM J03338+3320 during the 2015 superoutburst. \citet{kat16j0333} noted that stage A superhumps first appeared near the end of the precursor and survived in the faint stage, after which the stage A$-$B transition occurred around the maximum brightness of the main superoutburst, based on their careful analyses of superhump period changes and amplitudes of superhumps. Although the individual maxima of superhumps were unable to be measured, stage A superhumps in the faint stage were also detected in the Kepler light curve of V1504 Cyg \citep{osa14v1504cygv344lyrpaper3}.

In the case of QZ Vir, we detected a constant period of $P_{\rm sh}$ = 0.061181 d in the faint stage. This period is 1.1\% longer than that obtained between BJD 2457089.45$-$91.87. \citet{Pdot} noted that the stage A superhump period is 1.0$-$1.5 \% longer than the stage B superhump period in the same system. Although our observations were absent around the maximum brightness of the main superoutburst, the obtained $O - C$ diagram in figure \ref{o-c} indicates that a stage transition occurred around BJD 2457087 ($E$ ${\sim}$ 100). The presence of stage transitions is also supported by the variations of the superhump amplitudes. \citet{Pdot} reported that stage A superhumps have the largest amplitudes among three stages, and that a regrowth of superhump amplitudes is frequently seen in the stage B$-$C transition. In the middle panel of figure \ref{o-c}, the large amplitudes of superhumps are seen before BJD 2457087 ($E$ $<$ 100), and a regrowth of the superhump amplitudes is observed around BJD 2457092 ($E$ ${\sim}$ 190). In combination with the above results, we can reasonably conclude that the superhumps during the faint stage in QZ Vir are indeed the stage A superhumps.

Although the present observation is a rare case that stage A superhumps appear in the faint stage, the question ramains how rare this phenomenon is. As noted above, there are only three SU UMa-type dwarf novae that exhibit stage A superhumps in the faint stage between the precursor and the main superoutburst: PM J03338+3320, V1504 Cyg, and QZ Vir. This rareness is due to insufficient data of the faint stage, or intrinsic rareness of this phenomenon. This should be clarified in future observations.

\subsection{Mass ratio and evolutional status of QZ Vir}

As noted in the introduction, the mass ratio of QZ Vir is unclear, which obscures an evolutional status of the object. In recent years, \citet{kat13qfromstageA} have established a new method to estimate a mass ratio using the stage A superhump period. The method can extend to other systems including WZ Sge-type dwarf novae and AM CVn stars, if the stage A superhump period and the orbital period of the system are known \citep{iso16crboo}. Using a stage A superhump period during the 2014 superoutburst of QZ Vir, \citet{Pdot7} estimated a mass ratio to be 0.18(2). However, \citet{Pdot7} noted that this value should be improved in future observations, since the stage A superhump period was determined under a short coverage of observations. 

In the present observations, sufficient data during stage A superhumps enables us to upgrade a mass ratio of QZ Vir. The obtained stage A superhump period and the known orbital period indicate a newly introduced excess/deficiency rate of ${\epsilon}^{*}$ = 0.0386(7), where ${\epsilon}^{*}$ is written as ${\epsilon}^{*}$ = 1 - $P_{\rm orb}/P_{\rm sh}$. Using table 1 of \citet{kat13qfromstageA}, we obtain $q$ = 0.108(3). This value is very typical for SU UMa-type dwarf novae with short orbital periods, and significantly smaller than that derived by \citet{sha84tleo}. The high mass ratio obtained by \citet{sha84tleo} ($q$ ${\sim}$ 0.48) is possibly due to overestimation of the $K_{1}$ value, which led to a light mass of the white dwarf ($M_{1} {\sim}$ 0.40$M_{\odot}$). The obtained mass ratio, orbital period and the absence of the secondary star in the optical spectrum exclude the possibility that QZ Vir contains an evolved secondary, like a possible progenitor for AM CVn stars (\cite{pod03amcvn}; \cite{ish07CVIR}). This mass ratio also excludes the period bouncer of the system \citep{lit06j1035}. The present result indicates that QZ Vir is an SU UMa-type dwarf nova evolving toward the period minimum. 

\subsection{Time scale of stage A superhumps}

\citet{lub91SHa} and \citet{lub91SHb} suggested that the growth time of superhumps are inversely proportional to $q^{2}$. Observationally, it takes about a week for WZ Sge-type dwarf novae to exhibit superhumps \citep{kat15wzsge}. \citet{kat15wzsge} discussed a time scale of stage A superhumps, showing that systems with small mass ratios tend to have long time duration of stage A superhumps. This tendency is particularly seen in candidates for period bouncers (see figure 22 of \citet{kat15wzsge}). In the case of the 2015 superoutburst of QZ Vir, the duration of stage A superhumps was about 5 d (${\sim}$ 80 cycles), which is comparable to that of the candidates for period bouncers. As noted above, the mass ratio of QZ Vir obviously excludes the period bouncer of the system.

In order to explain such a long duration of stage A superhumps in the faint stage between the precursor and the main superoutburst of SU UMa-type dwarf novae, \citet{kat16j0333} suggest that a growth time of stage A superhumps needs a long time in a cold accretion disk. It is likely that the low viscosity state in the cold accretion disk requires more times to spread the eccentric mode into the entire disk \citep{kat16j0333}. The present observations further support this scenario. Based on the present and previous observations concerning the duration of stage A superhumps, it may be that the duration of stage A superhumps is mainly determined by two factors: mass ratio of the system and viscosity (or temperature) of the accretion disk.

\section{Summary}

We summarize the results of this paper as follows:

\begin{itemize}
\item We observed the 2015 February-March superoutburst of QZ Vir. The main superoutburst was separated from the precursor with the interval of ${\sim}$ 5 d. After the main superoutburst, QZ Vir showed a rebrightening.

\item Superhumps were observed after the end of the precursor. During the faint stage between the precursor and main superoutburst, we estimated the superhump period to be 0.061181(42) d. This period was ${\sim}$ 1.1\% longer than that observed in the previous superoutbursts. Based on the obtained $O - C$ diagram of the superhump maxima, the long period of the superhumps, and variations of the superhump amplitudes, we identified the superhumps in the faint stage as stage A superhumps.

\item Using the refined stage A superhump period, we determined a mass ratio of QZ Vir to be 0.108(3). In combination with the mass ratio and the optical spectrum of QZ Vir obtained by \citet{sha84tleo}, we exclude the possibility that QZ Vir contains an evolved secondary star. This value also rules out the period bouncer of the system. The present result indicates that QZ Vir is an SU UMa-type dwarf nova evolving toward the period minimum.

\item It is likely that a cold accretion disk lengthens the duration of stage A superhumps. In combination with the present observations and the previous reports regarding the duration of stage A superhumps, a time scale of stage A superhumps is determined by the temperatures of the accretion disk and mass ratio of the system.

\end{itemize}

\begin{ack}
We would like to thank the anonymous referee for helpful comments on the manuscript of the paper. This work was supported by the Grant-in-Aid Initiative for High-Dimensional Data-Driven Science through Deepening of Sparse Modeling E(25120007) from the Ministry of Education, Culture, Sports, Science and Technology (MEXT) of Japan. The authors are grateful to observers of VSNET Collaboration and VSOLJ observers. We acknowledge with thanks the variable star observations from the AAVSO International Database contributed by observers worldwide and used in this research.
\end{ack}

\begin{figure*}
\begin{center}
\includegraphics[width=8cm]{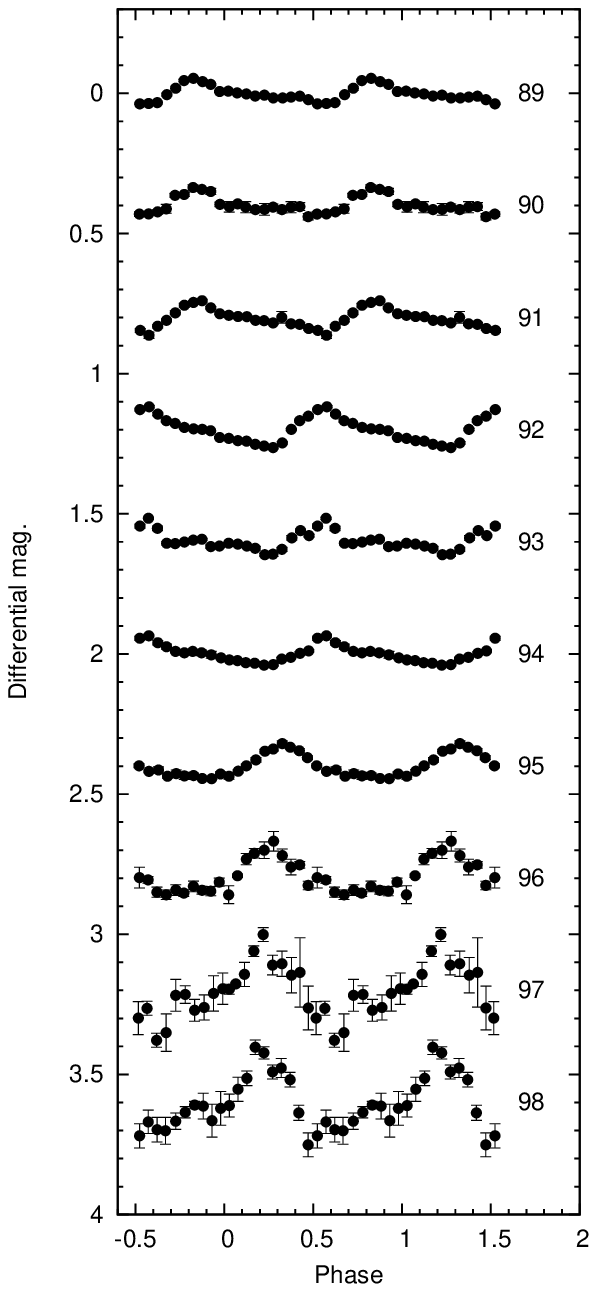}
\includegraphics[width=8cm]{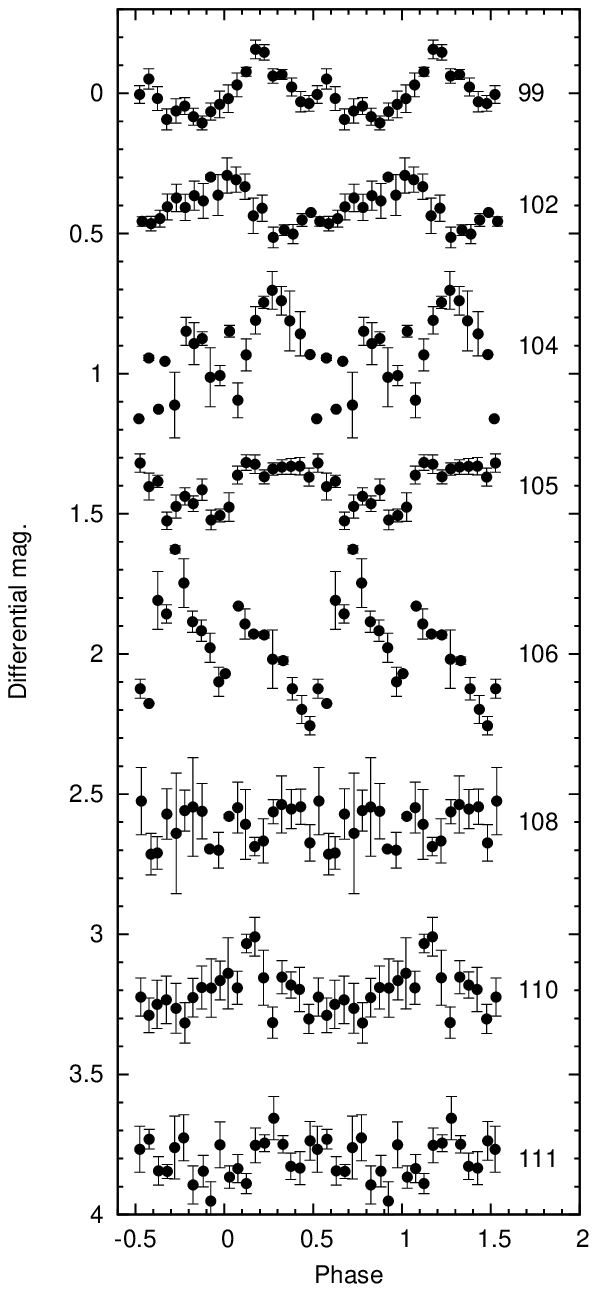}
\end{center}
\caption{Phase-averaged light curves between BJD 2457089$-$111. The numbers in this figure denote the day since BJD 2457000. The light curves are folded with 0.060490 d for BJD 2457089$-$91, 0.059944 d for BJD 2457092$-$93, and 0.060000 d for BJD 2457094$-$111, respectively.}
\label{shvar2}
\end{figure*}



\begin{longtable}{ccccc}
\caption{Times of superhump maxima.}\label{toc}
\hline
\hline
$E$$^{*}$ & maximum time$^{\dagger}$ & error$^{\ddagger}$ & $O - C$$^{\S}$ & $N$$^{||}$ \\
\endfirsthead
\hline
0 & 7080.50961 & 0.00064 & -0.07400 & 52 \\
52 & 7083.69374 & 0.00362 & -0.02028 & 16 \\
53 & 7083.76876 & 0.00393 & -0.00545 & 21 \\
54 & 7083.81603 & 0.00073 & -0.01838 & 24 \\
55 & 7083.87019 & 0.00116 & -0.02443 & 17 \\
68 & 7084.68263 & 0.00284 & 0.00542 & 12 \\
69 & 7084.73715 & 0.00228 & -0.00026 & 22 \\
70 & 7084.79637 & 0.00148 & -0.00124 & 23 \\
71 & 7084.85364 & 0.00133 & -0.00417 & 23 \\
85 & 7085.72615 & 0.00214 & 0.02554 & 22 \\
86 & 7085.77326 & 0.00242 & 0.01245 & 22 \\
87 & 7085.83802 & 0.00120 & 0.01701 & 22 \\
147 & 7089.45464 & 0.00040 & 0.02162 & 54 \\
148 & 7089.51574 & 0.00052 & 0.02252 & 58 \\
149 & 7089.57698 & 0.00043 & 0.02357 & 59 \\
168 & 7090.72464 & 0.00114 & 0.02743 & 23 \\
168 & 7090.66485 & 0.00146 & 0.02784 & 14 \\
170 & 7090.78725 & 0.00136 & 0.02984 & 23 \\
176 & 7091.14929 & 0.00105 & 0.03067 & 87 \\
177 & 7091.21183 & 0.00085 & 0.03302 & 84 \\
178 & 7091.27271 & 0.00118 & 0.03369 & 45 \\
179 & 7091.33209 & 0.00153 & 0.03288 & 26 \\
181 & 7091.45349 & 0.00058 & 0.03387 & 57 \\
182 & 7091.51290 & 0.00037 & 0.03309 & 62 \\
183 & 7091.57191 & 0.00027 & 0.03189 & 62 \\
187 & 7091.81342 & 0.00031 & 0.03261 & 41 \\
188 & 7091.87310 & 0.00039 & 0.03209 & 37 \\
193 & 7092.17421 & 0.00028 & 0.03220 & 68 \\
194 & 7092.23426 & 0.00026 & 0.03205 & 89 \\
195 & 7092.29478 & 0.00035 & 0.03236 & 84 \\
201 & 7092.65381 & 0.00058 & 0.03020 & 18 \\
202 & 7092.71323 & 0.00040 & 0.02942 & 33 \\
203 & 7092.83293 & 0.00050 & 0.02872 & 39 \\
203 & 7092.77338 & 0.00035 & 0.02936 & 38 \\
204 & 7092.89335 & 0.00079 & 0.02894 & 21 \\
208 & 7093.13265 & 0.00036 & 0.02743 & 136 \\
217 & 7093.67244 & 0.00227 & 0.02542 & 9 \\
219 & 7093.79362 & 0.00091 & 0.02620 & 15 \\
221 & 7093.85156 & 0.00106 & 0.02395 & 15 \\
225 & 7094.15175 & 0.00036 & 0.02314 & 90 \\
233 & 7094.63351 & 0.00072 & 0.02329 & 20 \\
234 & 7094.69314 & 0.00094 & 0.02273 & 17 \\
235 & 7094.75238 & 0.00075 & 0.02176 & 20 \\
236 & 7094.81149 & 0.00084 & 0.02068 & 19 \\
248 & 7095.53070 & 0.00068 & 0.01748 & 18 \\
249 & 7095.59071 & 0.00062 & 0.01729 & 26 \\
250 & 7095.65293 & 0.00128 & 0.01932 & 16 \\
251 & 7095.70905 & 0.00101 & 0.01523 & 18 \\
252 & 7095.77096 & 0.00150 & 0.01695 & 18 \\
253 & 7095.83194 & 0.00068 & 0.01773 & 19 \\
258 & 7096.12809 & 0.00079 & 0.01288 & 80 \\
263 & 7096.43063 & 0.00049 & 0.014413 & 51 \\
283 & 7097.62215 & 0.00166 & 0.00193 & 15 \\
284 & 7097.68121 & 0.00148 & 0.00079 & 14 \\
296 & 7098.39190 & 0.00099 & -0.01091 & 33 \\
297 & 7098.46929 & 0.00096 & 0.00627 & 30 \\
298 & 7098.52789 & 0.00113 & 0.00468 & 33 \\
313 & 7099.41961 & 0.00123 & -0.00660 & 33 \\
314 & 7099.48429 & 0.00081 & -0.00212 & 29 \\
315 & 7099.54617 & 0.00127 & -0.00045 & 33 \\
368 & 7102.71511 & 0.00139 & -0.02211 & 16 \\
369 & 7102.77378 & 0.00220 & -0.02364 & 16 \\
\hline
\multicolumn{3}{l}{$^*$ Cycle count. $^{\dagger}$ BJD-2450000.} \\
\multicolumn{3}{l}{$^{\ddagger}$ errors in units of days. $^{\S}$ against equation \ref{eq1}.} \\
\multicolumn{3}{l}{$^{||}$Number of datapoints to determine the maximum time.} \\
\end{longtable}


\bibliographystyle{pasjtest1}
\bibliography{cvs2016}

\end{document}